\documentclass{svmult}
\usepackage{times}
\usepackage{amsmath}
\usepackage{graphicx}        
\usepackage[bottom]{footmisc}
\usepackage{epstopdf}

\begin{document} 

\title{Energy-Aware Lease Scheduling in Virtualized Data Centers}
\author{Nguyen Quang-Hung, Nam Thoai, Nguyen Thanh Son, Duy-Khanh Le}

\institute{Faculty of Computer Science and Engineering, HCMC University of Technology, VNUHCM, \\
268 Ly Thuong Kiet Street, District 10, Ho Chi Minh City, Vietnam\\
\texttt{\{hungnq2, nam, sonsys\}@cse.hcmut.edu.vn}, \\
Department of Computer Science, National University of Singapore, \\
\texttt{leduykha@comp.nus.edu.sg}}

\maketitle
\vspace{-18mm}
\abstract{
Energy efficiency has become an important measurement of scheduling algorithms in virtualized data centers. 
One of the challenges of energy-efficient scheduling algorithms, however, is the trade-off between minimizing 
energy consumption and satisfying quality of service (e.g. performance, resource availability on time for reservation requests). 
We consider resource needs in the context of virtualized data centers of a private cloud system, which provides resource leases in terms of virtual machines (VMs) for user applications.
In this paper, we propose heuristics for scheduling VMs that address the above challenge. On performance evaluation, simulated results have shown a significant reduction on total energy consumption of our proposed algorithms compared with an existing First-Come-First-Serve (FCFS) scheduling algorithm with the same fulfillment of performance requirements. We also discuss the improvement of energy saving when additionally using migration policies to the above mentioned algorithms.
}

\vspace{-2mm}
\section{Introduction}
\label{sec:1}
Cloud computing \cite{Buyya2009a} has been developed as a utility computing model and is driven by economies of scale. 
Reduction in energy consumption (kWh) for cloud systems, which are built up from
virtualized data centers \cite{Sotomayor2010, Beloglazov2012}, is of
high concern for any cloud provider. Energy-aware scheduling of VMs in virtualized data centers is still challenging \cite{Albers2010, VonLaszewski2009, Goiri2010, Beloglazov2012}. 
There are several works that have been proposed to address the problem of energy-efficient scheduling of VMs in cloud data centers. Some works \cite{Albers2010, VonLaszewski2009} 
proposed scheduling algorithms to change adaptatively processor speed when executing user applications such that the changing processor speed method meets 
user requirements and reduces power consumption of processors when executing user applications. Some other works proposed algorithms that consolidate VMs 
a onto small set of physical servers in a virtualized datacenter \cite{Goiri2010, Beloglazov2012} such that power consumption of physical servers is minimized. 
However, the challenge on reducing energy consumption while preserving quality of service (e.g. performance or resource availability on time for reservation request) remains.

Sotomayor et al. \cite{ Sotomayor2008, Sotomayor2010} have proposed a lease-based model for the resource provisioning problems and presented 
FCFS-based scheduling 
algorithms to meet user performance. The presented scheduling algorithms in that works, however, have never involved energy efficiency. 
In this paper, we introduce an energy-aware lease scheduling problem with trade-off between minimizing of energy consumption and satisfying quality of service. 
We concern on the provision of hardware resources. The software requirements on provisioning resource are out of scope of this paper. 
Using VMs incurs some overheads (e.g. transferring VM images); therefore, these overheads of VMs should be considered in the problem of scheduling 
VM-based leases. 
The resource allocation problem of VMs with multiple resources is NP-hard. Each VM requires multiple resources such as CPU, memory, I/O
 to execute its applications. The resource allocation problem can be seen as a d-dimensional Vector Bin Packing problem ($VBP_{d} $) \cite{Panigrahy2011}, 
 in which each physical server with multiple resources is considered as a d-dimensional bin, and each virtual machine is a d-dimensional item with 
 various sizes of requested resources (e.g. CPU, memory). The $VBP_{d} $ is claimed as NP-hard problem for $ \forall d \geq 1 $ \cite{Panigrahy2011}.

In recent research, Fan et al. \cite{Fan2007a} claimed a linear relationship between power consumption (in Watts) on a physical server and its 
load (i.e., CPU utilization). The authors estimate that the power consumption of an idle (0\% CPU utilization) server is equal or greater than fifty percent
 of the power consumption of the server at a full load (100\% CPU utilization). Barroso and H\"{o}lzle \cite{Ittner2007} have proposed a case of 
 energy-proportional computing where all components in a computer
 could be turned on/off on demand. In this paper, we propose an energy-aware scheduling algorithm to map user lease requests onto physical servers. 
 The objective of our scheduling algorithm is to find an optimal schedule that has a minimum number of active physical servers and finishes all 
 user lease requests while satisfying user lease requirements. Our scheduling algorithm includes two phases: power-aware VM allocation and re-scheduling. 
 Our proposed allocation algorithm uses the minimum number of physical
 servers on mapping of the ready leases (in scheduler's queue). We
 also solve a re-scheduling problem by suspending, migrating, and resuming leases from physical servers that have CPU utilization lower than a 
 pre-defined low-threshold. These low load physical servers could be put into energy saving modes (e.g. stand-by, suspend to disk, or turn idle nodes off) 
 to avoid unwanted power consumption (e.g. fifty percent) in idle nodes \cite{Beloglazov2012}.

The remainder of the paper is organized as follows. In Section 2, we discuss the works that are related to 
our approach and energy-aware scheduling of virtual machines in virtualized data centers. We present the lease 
scheduling problem and the proposed energy-aware scheduling and
migration algorithms in Section 3. The results of our simulation study are reported and discussed in Section 4. The last section gives conclusions and future work.

\vspace{-2mm}
\section{Related works}
\label{sec:2}
Sotomayor et al. \cite{Sotomayor2008, Sotomayor2010} proposed a lease-based model and implemented 
First-Come-First-Serve (FCFS) \cite{Feitelson2004} and back-filling \cite{Feitelson2004} algorithms to schedule best effort, 
immediate and advanced reservation leases. The FCFS and back-filling
algorithms consider only one performance metric such as waiting time and slowdown, without mentioning energy efficiency. To maximize performance, these scheduling algorithms tend to choose free load servers (i.e. those with the highest-ranking scores) when allocating a new lease. Therefore, a lease with just a single VM can be allocated on a big, multi-core physical server. This could waste a lot of energy. The authors also proposed a migration algorithm for preempting a best-effort lease in case the scheduler needs more resources for an advanced reservation lease. However, the authors did not use the migration algorithm on dynamic consolidation of VMs to turn low utilization servers off for energy saving. Instead, our allocations will choose working physical servers and turn off other free load servers. We also improve the migration algorithm to allow migration of leases that are running on low utilization servers, and turn these servers off.

Albers et al. \cite{Albers2010} reviewed some energy-efficient algorithms which are used to minimize flow time by changing processor speed according to job size. Laszewski et al. \cite{VonLaszewski2009} proposed scheduling heuristics and presented application experience for reducing power consumption of parallel tasks in a cluster with the Dynamic Voltage Frequency Scaling (DVFS) technique. We did not use the DVFS technique to reduce energy consumption on data centers.

Previous research \cite{Goiri2010}, \cite{Beloglazov2012} presented
scheduling algorithms that place virtual machines (VMs) in virtualized
data centers to minimize energy consumption. Beloglazov et al.
\cite{Beloglazov2012} presented a modified best-fit decreasing
(denoted as MBFD) heuristic for placement of VMs and VM migration
policies under adaptive thresholds in virtualized data centers. The
MBFD sorts all VMs in a decreasing order of CPU demands and tends to
allocate a VM to an active physical server that would take the minimum
increase of power consumption. The MBFD can reduce energy consumption
in a heterogeneous environment.
On the other hand, choosing a host with least increasing power 
consumption can lead to performance inefficiency. 
The MBFD will prefer a lower-performance host rather than a higher-performance host if each processor
in the lower-performance host consumes less power than each processor in the higher-performance host does. 
The MBFD is also not concerned about the duration time of VMs.
In contrast, our proposed allocation algorithms account for the
duration time of VMs and will greedily allocate VMs belonging to
a lease to the same physical machine. 
The previous migration policies \cite{Beloglazov2012} did not concern on overheads of migration (e.g. suspend, resume, and migration time) of VMs. We study effects of the overheads of migration of VMs on a schedule plan.
An optimum allocation of each independent VM is studied in
\cite{Goiri2010}. In the paper, the authors developed a score-based
allocation method to calculate the scores matrix of allocations of $ m $ VMs to $ n $ physical servers. A score is sum of many factors such as power consumption, hardware and software fulfillment, resource requirement. These studies are unsuitable for the following lease scheduling in this paper. We consider the case where each user lease has a limited duration time and contains a group of concurrent VMs (e.g. each MPI job requires tens to thousands of VMs concurrently).

\vspace{-2mm}
\section{Problem Description}
\label{sec:3}

Given a set of leases $ L_{i} $ (i $\in$ [1;n]) to be scheduled on a
set of physical servers $ M_{j} $ (j $\in$ [1;m]). We extend the resource model that is defined in \cite{Sotomayor2010}. A user requests some leases. A user $i^{th}$ lease requests (1) a set of $ rn_{i} $ identical virtual machines (VMs), (2) start time ($st_{i} $), and (3) duration of the lease ($ dur_{i}$). In the user $i^{th}$ lease, each $k^{th}$ VM requires $ u_{ik} $ percent of CPU utilization (e.g. each 100\% is one core), $ r_{ik} $ MB of memory, $ d_{ik} $ MB of disk image, and $ b_{ik} $ MB/s of network bandwidth. A lease can be a best-effort or an advanced reservation lease that is without or with user specified start time. Each physical server has total $ U $ percent of CPU utilization, $ R $ megabytes (MB) of memory, $ D $ MB of available file system, $ Bw $ MB/s of network bandwidth.


In this paper, we use the following energy consumption model proposed in \cite{Fan2007a, Beloglazov2012}:
\begin{align}
 P_{j} = P_{idle} + (P_{max} - P_{idle}) \times CPU_{j} 
\end{align}

where $ P_{idle}, P_{max} $, and $ P_{j} $ are idle power, maximum power, and total system power of a single physical server ($M_{j}$), and $ CPU_{j} $ is the server’s CPU utilization where $ 0 \leq CPU_{j} \leq 1 $.

The objective is to find an optimal schedule that maps all user lease requests into the smallest number of physical servers in order to minimize total energy consumption of all activated physical machines and to satisfy QoS (e.g. performance, or resource is available on time for advanced reservation leases \cite{Sotomayor2010}). Formally, we formulate static VM allocation problem as following:

\begin{center}
$ \textbf{Minimize} \sum_{j=1}^{m} ( P_{idle} + (P_{max} - P_{idle}) \times CPU_{j} ) \times y_{j}  $
\end{center}

subject to 
\vspace{-2mm}
\begin{align}
\sum_{i=1}^{n}  \sum_{k=1}^{rn_{i}} u_{ik} x_{ikj} \leq U_{j} \times y_{j},  \qquad  j=1,..., m     \\    
\sum_{i=1}^{n} \sum_{k=1}^{rn_{i}} r_{ik} x_{ikj} \leq R_{j} \times y_{j},  \qquad  j=1,..., m	\\
\sum_{i=1}^{n} \sum_{k=1}^{rn_{i}} b_{ik} x_{ikj} \leq  Bw_{j} \times y_{j}, \qquad j=1,..., m	\\
\sum_{i=1}^{n} \sum_{k=1}^{rn_{i}} d_{ik} x_{ikj} \leq D_{j} \times y_{j}, \qquad j=1,..., m 	\\
\sum_{i=1}^{n} \sum_{j=1}^{m}  x_{ikj} = 1, \qquad  k=1,..., rn_{i} 
\end{align}

\begin{align}
CPU_{j} = \dfrac{ \sum_{i=1}^{n} \sum_{k=1}^{rn_{i}}  u_{ik} x_{ikj} }{U},  \qquad  j=1,..., m \\
x_{ikj} \leq y_{j}, \qquad  i=1,..., n, k=1,..., rn_{i}, j=1,..., m
\end{align}


{\noindent}where the binary variables $ x_{ikj} \in \{0, 1\} $ and $ y_{j} \in \{0, 1\} $. $ x_{ikj} = 1 $ if and only if the $k^{th}$ VM of the lease $ L_{i} $ is allocated on the $ M_{j} $, and $ y_{j} = 1 $ if and only if the $ M_{j} $ is allocating resources for at least one VM and $ y_{j} = 0 $ if and only if the  $ M_{j} $ is in a sleep state. (That is we assume that a server in sleep state does not consume energy). 
The equations (1.2) to (1.5) are constraints on resources of each physical server, the equation (1.6)  describes the fact that each VM will be allocated on only one physical machine. 
The CPU utilization of a physical machine is calculated by the equation (1.7). 
We assume that the CPU utilization is unchanged during an interval of two
continuous events of the scheduler. The energy consumption ($E_{j}$) of a physical machine in period of [0;T] formulates as:
\begin{align}
 E_{j} = \int_{0}^{T} P_{j}(t)dt 
\end{align}

The makespan of a schedule ($ C_{max} $), is defined as the maximum of completion time of all leases and formulated as:
$ C_{max} = max \{ C(L_{i}) | i=1,...,n \} $, where the $ C(L_{i}) $ is completion time of a lease $L_{i}$. 
The $C(L_{i})$  formulated as $C(L_{i}) = ( st_{i} + dur_{i} + t_{i}^{mig} +  t_{i}^{sus} +t_{i}^{trans} ) $, where $st_{i}$ , $dur_{i}$, $t_{i}^{mig}$ , $t_{i}^{sus}$, $ t_{i}^{trans} $ are start time, duration time, migration time, suspend time, and  transferring time of image-disks of some VMs of the lease respectively. 

\vspace{-10mm}
\subsection{A special case}

Given a set of leases $ L_{i} $ (i $\in$ [1;n]) to be scheduled on a
set of identical physical servers $ M_{j} $ (j $\in$ [1;m]). Let us
assume that all user leases request only one VM. We formulate the
special lease scheduling with a single-VM problem as following:
\begin{center}
$ \textbf{Minimize} \sum_{j=1}^{m} E_{0} \times T_{j} + \sum_{i=1}^{n} e_{i} $
\end{center}

where
 $ E_{0} $ is the base energy consumption of a physical server in a unit of time, 
 $ T_{j} $ is the working time of a physical server $ M_{j}$ (j $\in$ [1;m]), 
 $ e_{i} $ is the energy consumption for executing a user lease $ L_{i} $ (i $\in$ [1;n]). 

\vspace{-10mm}
\subsection{Scheduling algorithm}

Our lease scheduling problem is on-line scheduling. The scheduling
algorithm is triggered by an event of a new lease  or at a regular
interval. Firstly, the algorithm sorts the  list of leases (e.g.
best-effort leases, immediate leases, etc.) in a  
 queue that are ready to run in decreasing order by lease duration.
 A lease that has longest duration time will be mapped first. Secondly, the algorithm uses a heuristic (FF-MAP-H2L or FF-MAP-L2H) for 
 mapping leases onto physical servers in order to minimize the number of active physical servers. The two allocation algorithms, FF-MAP-H2L 
 and FF-MAP-L2H, which are discussed in our previous works
 \cite{Hung2011}, both use two ways in sorting the list of physical
 servers (i.e. in the order of highest to lowest ranking scores of physical servers and reverse). They allocate a new lease to some active physical servers such that every VM in the new lease is allocated successfully. They always sort free load physical servers at the tail of the sorted list of physical servers. Our energy-aware lease scheduling algorithm is presented in Algorithm 1. 

\begin{table}[h]
\centering
\begin{tabular}{p{11cm}}
\textbf{Algorithm 1:} Energy-aware lease scheduling \\
\hline \textbf{Input:} leases in queue, set of physical hosts \\
\textbf{Output:} None or a mapping of scheduled leases \\
1: $ Q = $ Sort ready leases in queue in decreasing order of their durations. \\
2: \textbf{For} each lease $ l $ in the sorted lease queue $ Q $ \\
3: \hspace{0.5cm} \hangindent=1cm Use \textbf{FF-MAP-H2L} or
\textbf{FF-MAP-L2H} to map the lease $ l $ to the first active physical server. \\
4: \textbf{End For} \\
5: \textbf{If} all leases in the queue are mapped successfully, return the mapping of scheduled leases.\\
6: \textbf{Else} return None. \\
\end{tabular} 
\end{table}


In this paper, we extend the FF-MAP-H2L with migration, called (i)
PMIG-LxHy-FF-MAP-H2L and (ii) MIG-LxHy-FF-MAP-H2L. Both of the two
algorithms (i) and (ii) do re-scheduling by migrating all of the running leases on physical servers $ M_{k}$ ($ k \in [1;m] $) that 
have resource utilization less than a defined low threshold ($ x $)
(e.g. 0.4) and medium threshold ($ y $) (e.g. 0.8). Then the scheduler sets the servers $ M_{k} $ passive and puts 
them in energy-saving mode (e.g. sleep, shut down). A system administrator sets our defined low and medium thresholds. 
The algorithm (i) differs from the algorithm (ii) by adding one more step to check whether there are enough available 
resources in set $ S_{med} $, where $ S_{med} = \{ M_{j} | \forall j \in [1;m] \wedge x < cpuload(h) \leq y \} $, or not 
before it re-schedules all of the running leases on low utilization servers. 

We also consider the overheads for migrating leases in both PMIG-LxHy-FF-MAP-H2L and MIG-LxHy-FF-MAP-H2L. Given a lease $ L_{i} $ with set of $ L_{iv} $ VMs, the overhead for migrating the lease $ L_{i} $ 
includes migration time $ t_{i}^{mig} $, $ t_{i}^{sus} $ suspend time
and $ t_{i}^{res} $ resume time of the set of the lease's VMs. The migration time includes $ t_{i}^{trans} $ transferring time of image-disks of these VMs. 
The scheduler can estimate the migration time, suspend and resume time before re-schedule the migrated leases in future. 
A. Beloglazov's work \cite{Beloglazov2012} did not consider the migration overheads.

For example, consider a lease with two (2) VMs where each VM requires 1024MB of physical memory, 4096MB of hard disk, a 
100MB/s network, and a physical memory bandwidth of 32MB/s. Then, we have: $ t_{i}^{sus} = t_{i}^{res} = 2 \times (1024/32) = 64.00 $ 
seconds, $ t_{i}^{mig} = 2 \times (4096/100) = 81.92 $ seconds. The
total migration time that is the sum of migration, suspend and 
resume times is 145.92 seconds. Consequently, the migration time causes the lease's waiting time increase.


\vspace{-2mm}
\section{Experimental study}

\begin{figure}
\centering
\includegraphics[width=10cm, height=6cm]{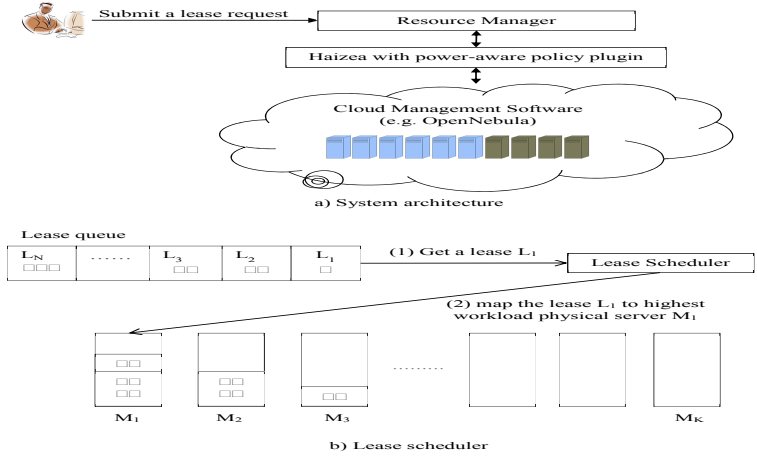}
\caption{The system architecture: (a) System architecture and (b) Lease scheduler}
\label{fig:1}
\end{figure}

The system architecture of an energy-efficient resource manager for private clouds was proposed in our previous work \cite{Hung2011}. Our proposed system has been deployed on a system with a cloud management software (e.g. OpenNebula) and a resource management (e.g. Haizea) in order to set up a private cloud.
Figure \ref{fig:1} shows the proposed system architecture (a) and lease scheduler (b) for provision resources.

We use a script, which is provided by Haizea \cite{Sotomayor2010}, to run and convert 30 days of a log trace in Parallel Archive Workload (SDSC-BLUE-2000-3.1-cln.swf \cite{SDSC}). We did not change information on the number of jobs, the job arrival time, time to finish the jobs during the conversion.
Each simulation will create a total of 5108 leases. Each lease has a various number of identical VMs with the same size (e.g. single core, 1024MB of RAM). 
We assume that the deployment of VMs on physical servers does not incur overheads. We assume that the simulated cloud data center has 1000 homogeneous physical servers. Each physical server has a 16/32-core CPU. Overheads of re-scheduling include the suspend/resume rate of 32MB/s and the network bandwidth of 100Mbps.


We experimented with the following lease allocation algorithms:

(1) Non Power-Aware Greedy (\textbf{NPA Greedy}): The original greedy algorithm in Haizea \cite{Sotomayor2010}.

(2-3) Our scheduling algorithm with \textbf{FF-MAP-L2H}, \textbf{FF-MAP-H2L}.

(4-6) The PMIG-LxHy-FF-MAP-H2L with three settings at 0.5, 0.4 and 0.3 low-threshold values and 0.8 high-threshold value that are denoted as \textbf{PMIG-L50H80-FF-MAP-H2L, PMIG-L40H80-FF-MAP-H2L and PMIG-L30H80-FF-MAP-H2L}.

(7-9) \textbf{MIG-L50H80-FF-MAP-H2L, MIG-L40H80-FF-MAP-H2L and MIG-L30H80-FF-MAP-H2L}: Running the MIG-LxHy-FF-MAP-H2L with three settings at 0.5, 0.4 and 0.3 low-threshold values and 0.8 high-threshold value

\begin{table}
\centering
\caption{Power consumption (Watt) of two HP Proliant servers (source from \cite{Spec2, Spec3})}
\label{tab:1} 
\begin{tabular}{|l|c|c|}
\hline Platform & $ P_{idle} $ & $ P_{max} $ \\
\hline HP Proliant DL585 G5 (2.7GHz, AMD Opteron 8384)	& 299\ W	& 521\ W \\
\hline HP Proliant DL785 G5 (2.30GHz, AMD Opteron 8376 HE)	& 444\ W	& 799\ W \\
\hline 
\end{tabular}
\end{table}

\vspace{-5mm}
\begin{figure}
\centering
\includegraphics[width=11cm, height=6cm]{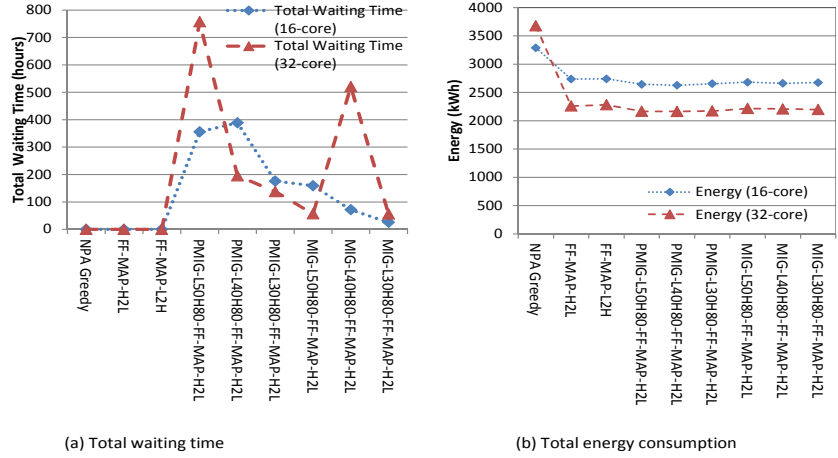}
\caption{The total energy consumption (kWh) for the investigated algorithms}
\label{fig:2}
\end{figure}

\vspace{-5mm}
\begin{table}[h]
\centering
\caption{Total energy consumption (kWh), total waiting time, and makespan ($ C_{max}) $ of lease allocation algorithms. Each server has 16 cores and 
16 GB of physical memory and the power model of HP Proliant DL585 G5 ($ P_{min} = 299Watts $, $ P_{max} = 521Watts $), $ T_{suspend} = T_{resume} = 32MB/s$, network bandwidth is 100Mbps.}
\label{tab:2} 
\begin{tabular}{|l|c|c|c|c|}
\hline Algorithm & Energy & Total waiting time & $ C_{max}$ & Total migrated leases\\
 & (kWh) & (hours) & (hours) 	 			& \\
\hline (1) NPA Greedy 				& 3287.59 & 0.000 	& 735.757	&  0 \\
\hline (2) FF-MAP-H2L				& 2736.07 & 0.000 	& 735.757	& 0 \\
\hline (3) FF-MAP-L2H				& 2741.61 & 0.000	& 735.757	& 0 \\
\hline (4) PMIG-L50H80-FF-MAP-H2L 	& 2644.36 & 355.869	& 737.246 & 483 \\
\hline (5) PMIG-L40H80-FF-MAP-H2L	& 2625.84 & 222.711 & 735.828 & 300 \\
\hline (6) PMIG-L30H80-FF-MAP-H2L	& 2654.22 & 175.804	& 736.943 & 223 \\
\hline (7) MIG-L50H80-FF-MAP-H2L 	& 2682.05 & 158.893	& 735.757 & 134 \\
\hline (8) MIG-L40H80-FF-MAP-H2L	& 2660.86 & 71.347 	& 735.757 & 165 \\
\hline (9) MIG-L30H80-FF-MAP-H2L	& 2674.44 & 25.438 	& 735.757 & 112 \\
\hline 
\end{tabular}
\end{table}

\begin{table}
\centering
\caption{Total energy consumption (kWh), total waiting time, $ C_{max} $ of lease allocation policies. Each server has 32 cores, 32 GB of physical memory and 
the power model of HP Proliant DL785 G5 ($ P_{min} = 444Watts $, $ P_{max} = 799Watts $), $ T_{suspend} = T_{resume} = 32MB/s$, network bandwidth is 100Mbps.}
\label{tab:3} 
\begin{tabular}{|l|c|c|c|c|}
\hline Algorithm & Energy & Total waiting time & $ C_{max}$ & Total migrated leases\\
 & (kWh) & (hours) & (hours) 	 			& \\
\hline (1) NPA Greedy				& 3676.35 & 0.000	& 735.757 &  0 \\
\hline (2) FF-MAP-H2L				& 2260.60 & 0.000 	& 735.757 &  0 \\
\hline (3) FF-MAP-L2H				& 2282.37 & 0.000 	& 735.757 &  0 \\
\hline (4) PMIG-L50H80-FF-MAP-H2L 	& 2165.67 & 757.395	& 736.943 &  464 \\
\hline (5) PMIG-L40H80-FF-MAP-H2L 	& 2167.33 & 195.388	& 736.989 &  297 \\
\hline (6) PMIG-L30H80-FF-MAP-H2L 	& 2171.52 & 137.541	& 735.828 & 225 \\
\hline (7) MIG-L50H80-FF-MAP-H2L 	& 2215.98 & 56.566  & 735.757 & 109 \\
\hline (8) MIG-L40H80-FF-MAP-H2L 	& 2207.44 & 520.333 & 735.757 &  113 \\
\hline (9) MIG-L30H80-FF-MAP-H2L 	& 2197.66 & 55.699  & 735.757 &  118 \\
\hline 
\end{tabular}
\end{table}

We collect experimental data on two physical server models: (i) HP Proliant DL585 G5 (2.7GHz, AMD Opteron 8384, 16GB of physical 
memory) \cite{Spec2}; and (ii) HP Proliant DL785 G5 (2.30GHz, AMD Opteron 8376 HE, 32GB of physical memory) \cite{Spec3}. Table \ref{tab:1} 
shows the average active power of both server models. Table \ref{tab:2} and Table \ref{tab:3} show simulation results of the above lease allocation 
algorithms on a simulated cluster with 16 and 32 core architectures and compare their total energy consumption (kWh) to the NPA Greedy 
algorithm \cite{Sotomayor2010}. Figure \ref{fig:2} shows the total energy consumption (kWh) of each allocation algorithm. 

The results show that the energy-aware lease scheduling has the total waiting time and $ C_{max} $ equal  to that of the NPA 
in the experiments. Compared to the NPA, the energy-aware lease scheduling with both FF-MAP-H2L and FF-MAP-L2H reduces 
the total energy consumption in both 16-core and 32-core cases. Our proposed algorithms reduced total energy consumption that is 
linear increasing in the number of cores in each host. Moreover, using the FF-MAP-H2L with migration algorithms at three (0.5, 0.4, 0.3) 
threshold values, called PMIG-L50H80-FF-MAP-H2L, PMIG-L40H80-FF-MAP-H2L, PMIG-L30H80-FF-MAP-H2L, MIG-L50H80-FF-MAP-H2L, 
MIG-L40H80-FF-MAP-H2L and MIG-L30H80-FF-MAP-H2L, also reduced the total energy consumption more than the FF-MAP-H2L, FF-MAP-L2H 
and NPA without migration. A disadvantage of these migration algorithms, however, is the decreasing performance, i.e. these migration 
algorithms increase the total waiting time of migrated leases when we consider overheads in migration and rescheduling these migrated leases. 
Consequently, $ C_{max} $ can be increased.


\vspace{-2mm}
\section{Conclusions and future work}
This work presents an energy-aware lease scheduling problem and
proposes a scheduling algorithm for lease scheduling problems to
minimize the total energy consumption. The simulation results show 
that our algorithms reduce the total energy consumption significantly
compared with an existing FCFS-based algorithm in the Haizea. Our algorithms are also beneficial on multi-core architectures, i.e. the more cores the 
machines have, the more the energy consumption is reduced. 

In future, we are interested in cloud systems with heterogeneous resources. The cloud systems will provide resources to many 
types of leases such as best-effort, advanced reservation, and
immediate leases at the same time. We will investigate the VM placement problem with multiple resources (e.g. CPU, RAM, network bandwidth, etc.) 
and scheduling algorithms to solve the special case of energy-aware lease scheduling.


\end{document}